# The $\text{sech}(\hat{\xi})$-type profiles: a Swiss-Army knife for exact analytical modelling of thermal diffusion and wave propagation in graded media

**J.-C. Krapez**

ONERA, The French Aerospace Lab, DOTA, F-13661 Salon de Provence, France

E-mail: jean-claude.krapez@onera.fr

**Abstract**

This work deals with exact analytical modelling of transfer phenomena in heterogeneous materials exhibiting one-dimensional continuous variations of their properties. Regarding heat transfer, it has recently been shown that by applying a Liouville transformation and multiple Darboux transformations, infinite sequences of solvable profiles of thermal effusivity can be constructed together with the associated temperature (exact) solutions, all in closed-form expressions (vs. the diffusion-time variable and with a growing number of parameters). In addition, a particular class of profiles, so-called $\text{sech}(\hat{\xi})$-type profiles, exhibit high agility and in the same time parsimony. In this paper we go further into the description of these solvable profiles and their properties. Most importantly, their quadrupole formulation is provided which allows building smooth synthetic profiles of effusivity of arbitrary complexity and thereafter getting very easily the corresponding temperature dynamic response. Examples are given with increasing variability of effusivity and increasing number of elementary profiles. These highly flexible profiles are equally relevant for providing an exact analytical solution to wave propagation problems in 1D graded media (i.e. Maxwell's equations, acoustic equation, telegraph equation...). From now on, let it be for diffusion-like or wave-like problems, when the leading properties present (possibly piecewise-) continuously heterogeneous profiles, the classical staircase model can be advantageously replaced by a "high-level" quadrupole model consisting of one or more $\text{sech}(\hat{\xi})$-type profiles, which makes the latter a true Swiss-Army knife for analytical modelling.

# 1 Introduction

Graded media are ubiquitous. Examples can be found at different scales, both in nature and in manmade materials. Living tissues (skin, teeth), wood, soils, rocks and atmosphere boundary layer are common examples. In manmade materials, the gradient structure may appear incidentally (as due, by instance, to matter diffusion), whereas in functionally graded materials (FGM) it is introduced on purpose for reaching target properties. In first instance, the gradient affects composition, microstructure or porosity [1]; some examples are: fibre reinforced layered composites, ceramic thermal barriers, case-hardened steels, armour plates.

When studying diffusion processes (e.g. heat diffusion) inside continuously heterogeneous media, difficulties appear since, even for the one-dimensional (1D) case, an exact general (closed-form) solution is lacking for *arbitrary profiles* of the leading parameters. One remedy consists in approximating these profiles; another one consists in approximating the transfer equation. In the first case, when the profiles are not too complex, one can try approaching them with profiles leading to known temperature analytical solutions. Examples are linear [2], power-law [2, 3], exponential [3-5] profiles of the conductivity or the specific heat, as well as trigonometric and hyperbolic profiles of the square root of conductivity [6]. In specifically transformed spaces, exponential [7, 8] and power law [9] profiles of effusivity also yield explicit closed-form solutions. The disadvantage for many of them is that they require the use of special functions (e.g. Bessel, Airy, hypergeometric functions...), which penalizes the computation time, especially when the temperature data have to be calculated iteratively, as for example in the core of an inversion routine. Profiles presenting more complex shapes require implementing other strategies, such as an approximation with staircase functions [10, 11] or with piecewise-linear functions [5, 12-15]. Obviously, the finer the spatial discretisation, the better the results [12, 15].

The second approach (i.e. approximation of the heat equation) consists in applying a perturbation method [16], a linearization of the non-linear differential equation describing the variations of the thermal reflection coefficient [17-18] or thermal impedance [19], the WKBJ method [20], or an empirical method [21] derived from the THO theory whose result is closely related to the latter. All these methods require that the thermal properties have small and slow variations with position.

A third group of methods is based on a series expansion of the solution: the Generalized Integral Transform method [22], the Spectral Parameter Power Series method [23] and an asymptotic expansion for thermal waves at high frequency [24]. All may admit arbitrary profiles of the thermal properties as input data (provided some integrability or derivability criteria are met), however, since the temperature solution involves infinite series, which unavoidably have to be truncated, these methods should also be considered approximate (in addition, the required number of terms in the series depends on the steepness of the profiles).

A method was described a few years ago for building sequences of *solvable profiles* for several types of transfer equations in graded materials, in particular the heat equation and the wave equation [25] (a *solvable profile* means a profile of the leading property for which an exact analytical expression of the thermal field or wave field solution can be found). The underlying technique is the Bäcklund transformation applied in the time domain. Recently we presented a simpler method, the PROFIDT method (PROperty and FIeld Darboux Transformation) which is equally efficient for diffusion equations (like the heat equation) [26, 27] and for wave equations (like Maxwell's equations describing electromagnetic (EM) wave propagation in lossless and source-free media with graded permittivity and permeability) [28]. In both cases, it should be applied in the Laplace domain or Fourier domain (in these domains, all solutions presented in [26-28] are in closed-form; some are in closed-form even after getting back in the time domain [27]). The PROFIDT method can be iterated to produce increasingly sophisticated solvable profiles of the leading property (namely thermal effusivity for the thermal problem and the EM tilted admittance for the EM-wave problem) together with the field solution (temperature or heat flux field, resp. electric or magnetic field). Another option is to use intermediate-order solvable profiles (that is, obtained after applying one single PROFIDT) and merge them; the dynamic response of the whole heterogeneous structure is then obtained by applying the transfer matrix method and multiplying the quadrupoles related to each elementary profile, as for any passive multilayer [29, 30]. A family of profiles has proved particularly interesting for this purpose; they were called "$\text{sech}(\hat{\xi})$-type" profiles. They are defined with four parameters and those can be adjusted so that the values of the modelled property at both ends of a given layer together with its first derivatives reach any set of four specified (finite) values. By taking advantage of this flexibility it was envisioned that smooth solvable profiles of any shape could be synthesized by assembling elementary profiles of $\text{sech}(\hat{\xi})$-type.

In this work, we aim to build upon the previous study and provide a thorough investigation and exploration on the properties of this special class of solvable profiles. Our goal is to advance the state of the art by developing effective analytical tools for modelling diffusion processes as well as wave propagation in complex media described by a 1D continuously heterogeneous distribution of their properties.

To present the approach and outcomes, this paper is organized as follows. In Sec 2, the main principles of the PROFIDT method are briefly recalled. In Sec 3, the $\text{sech}(\hat{\xi})$-type profile definition is reported followed by a presentation of their properties. In Sec 4, we describe the tools necessary for applying the quadrupole methodology in the context of $\text{sech}(\hat{\xi})$-type heterogeneity. Examples with gradual complexity and relevant to modulated photothermal experiments are presented in Sec 5. Finally, Sec 6 is a discussion on applications to other kinds of problems, like wave propagation in media with continuously heterogeneous wave celerity, in particular light transmission in dielectrics with graded refractive index.

## 2 Fundamentals of the PROFIDT method [26]

Let us consider 1D heat diffusion in a medium with graded thermal conductivity $\lambda(z)$ and graded volumetric heat capacity $c(z)$. A Laplace transformation (or a Fourier transformation) is first applied which changes temperature $T(z,t)$ into $\theta(z,p)$ and heat flux density $\varphi(z,t)$ into $\phi(z,p)$ where the variable $p$ stands for either the Laplace variable (transient regime) or the Fourier variable $i\omega$ where $\omega$ is the angular frequency $2\pi f$ (periodic regime). The heat equation is usually expressed in terms of temperature (we will then use the symbol $\langle T \rangle$):

$$\langle T \rangle: \qquad p\theta = \frac{1}{c(z)} \frac{d}{dz}\left(\lambda(z)\frac{d\theta}{dz}\right) \tag{1}$$

A similar equation can be found for the heat flux density $\phi = -\lambda \theta_z$ where, remarkably enough, $\lambda(z)$ and $c^{-1}(z)$ are simply interchanged with respect to the previous one (the symbol $\langle \varphi \rangle$ will be used for all features related to this second equation):

$$\langle \varphi \rangle: \qquad p\phi = \lambda(z)\frac{d}{dz}\left(\frac{1}{c(z)}\frac{d\phi}{dz}\right) \tag{2}$$

A Liouville transformation is then applied. The first step consists in changing the *independent* variable $z$ (i.e. the physical depth) into the *square root of diffusion time* (SRDT) $\xi$, which is defined from the diffusivity profile $a(z) = \lambda(z)/c(z)$ according to (the same origin is taken for both $z$-scale and $\xi$-scale):

$$\xi = \int_0^z a^{-1/2}(u)du \tag{3}$$

The second step consists in changing the *dependent* variable, either the temperature (in eq. (1)) or the heat flux (in eq. (2)), into a new one (invariably written $\psi(\xi,p)$) by multiplying the former by a function $s(\xi)$ that is related to the thermal effusivity profile $b(z) = \sqrt{\lambda(z)c(z)}$ in two different ways depending on the considered form, $\langle T \rangle$ or $\langle \varphi \rangle$:

$$\langle T \rangle: \qquad \theta(\xi,p) \to \psi(\xi,p) \equiv \theta(\xi,p)s(\xi); \quad s(\xi) \equiv b^{+1/2}(\xi) \tag{4}$$

$$\langle \varphi \rangle: \qquad \phi(\xi,p) \to \psi(\xi,p) \equiv \phi(\xi,p)s(\xi); \quad s(\xi) \equiv b^{-1/2}(\xi) \tag{5}$$

In both cases, the new dependent variable $\psi(\xi, p)$ satisfies a stationary Schrödinger equation in which the ratio $s''(\xi)/s(\xi)$ (from now on, a prime denotes a derivation with respect to $\xi$) plays the role of the "potential" function, let us call it $V(\xi)$. First, this tells us the paramount importance of the effusivity profile when modelling 1D dynamic heat transfer in graded media. Next, this particular form of potential has the very interesting consequence that each of the former Schrödinger equations can be split into a system of two homologous Schrödinger equations, which means equations sharing the same potential function $V(\xi)$. The first one, with the variable $p$ acting as an eigenvalue, is for the "field" function $\psi(\xi, p)$; the second one, with a vanishing eigenvalue, is for the "property" function $s(\xi)$:

$$\begin{cases} \psi'' = (V(\xi) + p)\psi & \text{(6a)} \\ s'' = V(\xi)s & \text{(6b)} \end{cases}$$

The problem comes down to seeking solvable profiles $V(\xi)$ for the general Schrödinger equation (6a) with arbitrary value of the constant parameter $p$. Obviously, in addition to solutions for $\psi(\xi, p)$ in Eq. (6a), this would simultaneously provide solutions for $s(\xi)$ in Eq. (6b) as a particular case of Eq. (6a) with $p=0$. Then, by using in the reverse order Eq. (4) or (5) (as appropriate), one concomitantly gets the temperature field (or the heat flux field) and the effusivity profile that is associated with, all in generic forms.

For start, simple analytical solutions for $b^{+1/2}(\xi)$ or $b^{-1/2}(\xi)$ can be obtained by choosing a constant value for the potential function $V(\xi)$. Positive (or negative) values yield linear combinations (LC) of hyperbolic (resp. trigonometric) functions of a scaled SRDT. On the other side, a nil potential gives rise to linear functions of $\xi$.

The Darboux transformation is a perfect method for enriching the previous solutions. This is a differential transformation which, given the general solution of Eq. (6a) with a solvable profile $V_0(\xi)$ and one particular solution, provides a new solvable profile $V_1(\xi)$ and the general solution of Eq. (6a) with $V(\xi) = V_1(\xi)$. It was the first time, in [26], for heat-like equations and in [28], for wave-like equations, that the Darboux transformation was applied in tandem for finding out new solvable profiles of the leading property (effusivity in the first case, EM admittance in the second case) and the related analytical closed-form expression of the diffusion or wave fields. The joint procedure was dubbed the PROFIDT method (PROperty and FIeld Darboux Transformation method). In addition, the PROFIDT method can be applied iteratively for getting more and more sophisticated solutions [26, 28]. As a matter of fact, at each step, the new set of effusivity profiles is enriched with up to two parameters.

We will now concentrate on a specific class of solvable profiles, the so-called $\text{sech}(\hat{\xi})$-type profiles. These have been generated by a single Darboux transformation, starting from a constant positive potential function.

## 3 The $\text{sech}(\hat{\xi})$–type profiles

3.1 Description in the (square-root) diffusion-time space (i.e. $\xi$-space)

They are defined as a linear combination of two linearly independent solutions of Eq. (6b), $B(\xi)$ and $D(\xi)$, with the potential $V(\xi) = \xi_c^{-2}(1 - 2\text{sech}(\hat{\xi}))$:

$$\begin{cases} B(\xi) = \text{sech}(\hat{\xi}) \\ D(\xi) = \sinh(\hat{\xi}) + \hat{\xi}\,\text{sech}(\hat{\xi}) \end{cases} \tag{7}$$

The function sech is the inverse of the hyperbolic cosine: $\text{sech} = 1/\cosh$. $\hat{\xi}$ is the result of a linear transformation of the SRDT, $\xi$, namely $\hat{\xi} = \xi/\xi_c + \tau$. It involves two free-parameters: $\xi_c$ (a characteristic SRDT) and $\tau$ that may take values in $]0,+\infty[$, resp. in $]-\infty,+\infty[$. With $A_B$ and $A_D$, two other free parameters, the effusivity profiles read:

$$b^{\pm 1/2}(\xi) = s(\xi) = A_B B(\xi) + A_D D(\xi) \tag{8}$$

The "$+1/2$" exponent in Eq. (8) refers to profiles originating from the $\langle T \rangle$-form heat equation and "$-1/2$" to those originating from the $\langle \varphi \rangle$-form equation. Two sub-classes can thus be generated, according to the exponent sign. The effusivity profiles in Eq. (8) are defined with *four* adjustable parameters: $\xi_c$, $\tau$ and the two multiplicative factors $A_B$ and $A_D$. It was shown in [26] that the profiles generated from the LC in Eq. (8) are highly flexible functions: they may be concave, convex or wavy and they exhibit arbitrary steepness at the layer boundaries. Indeed, over a given $\xi$-interval, any *four* boundary conditions set can be satisfied, namely two end-values for $s(\xi)$ (non-zero and finite values) and two others for its derivative (finite values). At the same time, they exhibit maximum parsimony in satisfying theses conditions since they are defined with *no more than four* parameters. For these reasons, the "$\text{sech}(\hat{\xi})$-type" profiles are expected to be quite efficient for fitting any target profile, if needed, by assembling two or more of them.

Simultaneously with the $s(\xi)$ function in Eq. (8), the PROFIDT method provided the associated field function $\psi(\xi,p)$ through a LC of two functions $K(\xi,p)$ and $P(\xi,p)$:

$$\psi(\xi,p) = A_K K(\xi,p) + A_P P(\xi,p) \tag{9}$$

which are defined by:

$$\begin{cases} K(\xi,p) \equiv \alpha S(\xi,p) - \sigma C(\xi,p) \\ P(\xi,p) \equiv \alpha C(\xi,p) - \sigma S(\xi,p) \end{cases} \tag{10}$$

with:

$$\begin{cases} C(\xi,p) \equiv \cosh(\alpha\xi) \\ S(\xi,p) \equiv \sinh(\alpha\xi) \end{cases} ; \quad \begin{cases} \sigma = \sigma(\xi) \equiv \xi_c^{-1} \tanh(\hat{\xi}) \\ \alpha = \alpha(p) \equiv \sqrt{p + \xi_c^{-2}} \end{cases} \qquad (11)$$

3.2 Back to the (real) depth-space (i.e. $z$-space)

The PROFIDT method yields effusivity (solvable) profiles and temperature analytical solutions that are defined in Liouville space, i.e. $\xi$-space. In some circumstances, it may be necessary to have a description of them in $z$-space as well. The simplest option is when the diffusivity profile $a(z)$ is known; this implies that $\xi$ is substituted in the expression of effusivity or temperature by $\xi(z)$ which is calculated from Eq. (3) by incorporating the expression of $a(z)$. The trivial case of a constant diffusivity ($a(z) = a_=$) leads to $\xi(z) = z/\sqrt{a_=}$. Other options were described in [26]; one of them yields an *implicit* description of the effusivity profile through the coupled relations:

$$\begin{cases} b = b(\xi) \\ z = z(\xi) \end{cases} \qquad (12)$$

where $z(\xi)$ is obtained by inverting Eq. (3), which amounts to a Liouville *inverse transformation*. For being effective, this inversion requires extra information on diffusivity, volumetric heat capacity or conductivity. Knowing that the inverse of Eq. (3) may take the following equivalent forms:

$$z = z(\xi) = \int_0^\xi \sqrt{a(u)}\,du = \int_0^\xi \frac{b(u)}{c(u)}\,du = \int_0^\xi \frac{\lambda(u)}{b(u)}\,du \qquad (13)$$

the extra information that either volumetric heat capacity or conductivity is constant (i.e. $c(z) = c_=$ or $\lambda(z) = \lambda_=$) allows simplifying the latter two quadratures. A constant volumetric heat capacity (this hypothesis is generally adopted when dealing with condensed phases, as for example case hardened steel) would imply calculating the primitive of effusivity in $\xi$-space. A constant conductivity (a situation that is less often met) would imply calculating the primitive of its inverse. Analytical expressions of these primitives were proposed in [26] for $\mathrm{sech}(\hat{\xi})$-type profiles, which can be summarized in a set of four equations:

$$z(\xi) = \frac{\xi_c}{c_=}\left[f(\hat{\xi}(\xi)) - f(\hat{\xi}(0))\right] \quad \text{for } c(z) = c_= \qquad (14)$$

$$z(\xi) = \xi_c \lambda_= \left[f(\hat{\xi}(\xi)) - f(\hat{\xi}(0))\right] \quad \text{for } \lambda(z) = \lambda_= \qquad (15)$$

$$f(\hat{\xi}) = A_B^{\,2} \tanh(\hat{\xi}) + A_D^{\,2}\left[\frac{1}{4}\sinh(2\hat{\xi}) - \frac{\hat{\xi}}{2} + \hat{\xi}^2 \tanh(\hat{\xi})\right] + 2 A_B A_D \hat{\xi} \tanh(\hat{\xi}) \qquad (16)$$

$$f(\hat{\xi}) = \begin{cases} \dfrac{-1}{2 A_D} \dfrac{\mathrm{sech}(\hat{\xi})}{b^{\pm 1/2}(\hat{\xi})} & \text{if } A_D \neq 0 \\[2mm] \dfrac{1}{2 A_B} \dfrac{\sinh(\hat{\xi}) + \hat{\xi}\,\mathrm{sech}(\hat{\xi})}{b^{\pm 1/2}(\hat{\xi})} & \text{if } A_B \neq 0. \end{cases} \qquad (17)$$

Eq. (16) should be used in the case of $\langle T \rangle$-form profiles and constant heat capacity (with Eq. (14)) or in the case of $\langle \varphi \rangle$-form profiles and constant conductivity (with Eq. (15)). On the other side, Eq. (17) should be used in the case of $\langle T \rangle$-form profiles (by selecting the positive exponent for effusivity at the denominator) and constant conductivity (with Eq. (15)) or in the case of $\langle \varphi \rangle$-form profiles (by selecting the negative exponent) and constant heat capacity (with Eq. (14)).

Furthermore, on can notice that Eq. (13) could also be written in the following form:

$$z = \int_0^{\xi} \lambda^q(u) c^{q-1}(u) b^{1-2q}(u) du \tag{18}$$

where $q$ is a constant real number. If it happened that $\lambda^q(z) c^{q-1}(z)$ be constant for a particular value of $q$, Eq. (18) would amount to calculate the primitive of $b^{1-2q}(\xi)$. Except for very special cases ($q=0$, $q=1/2$, and $q=1$ correspond to the previous cases of constant $c(z)$, constant $a(z)$ and constant $\lambda(z)$) this would require a numerical computation of this quadrature.

## 4 Tools for implementing the quadrupole methodology

Useful relations regarding both $\langle T \rangle$-form and $\langle \varphi \rangle$-form profiles are first recalled [26]:

$$\langle T \rangle: \begin{cases} b = s^2 \\ \theta = \psi \, s^{-1} \\ \phi = -W(s, \psi) \end{cases} \qquad \langle \varphi \rangle: \begin{cases} b = s^{-2} \\ \theta = -p^{-1} W(s, \psi) \\ \phi = \psi \, s^{-1} \end{cases} \tag{19}$$

where $W$ is the Wronskian: $W(s, \psi) = s \psi' - s' \psi$.

The quadrupole (or transfer) matrix $\mathbf{M}$ of a layer extending from $z = z_0$ to $z = z_1$ (i.e. from $\xi = \xi_0$ to $\xi = \xi_1$ in $\xi$-space; in the sequel, 0 or 1 index means that the functions is evaluated at $\xi_0$, resp. at $\xi_1$) relates the input temperature/heat flux vector, i.e. $[\theta_0 \ \phi_0]^t$ to the output vector, i.e. $[\theta_1 \ \phi_1]^t$ as follows (see e.g. [29, 30]):

$$\begin{bmatrix} \theta_0 \\ \phi_0 \end{bmatrix} = \mathbf{M} \cdot \begin{bmatrix} \theta_1 \\ \phi_1 \end{bmatrix} \tag{20}$$

Quadrupoles $\mathbf{M}_{\langle T \rangle}$ and $\mathbf{M}_{\langle \varphi \rangle}$ are obtained when considering profiles of $\langle T \rangle$-form, resp. of $\langle \varphi \rangle$-form. Their calculation follows the classical procedure described, for instance, in [29, 30]. Hence, a synthetic expression for $\mathbf{M}_{\langle T \rangle}$ is given by:

$$\mathbf{M}_{\langle T \rangle} = \begin{bmatrix} K/s & P/s \\ -W(s, K) & -W(s, P) \end{bmatrix}_0 \times \begin{bmatrix} K/s & P/s \\ -W(s, K) & -W(s, P) \end{bmatrix}_1^{-1} \tag{21}$$

After some algebra, the four terms of the matrix are expressed as follows:

$$\begin{bmatrix} A_{\langle T \rangle} & B_{\langle T \rangle} \\ C_{\langle T \rangle} & D_{\langle T \rangle} \end{bmatrix} = \frac{1}{\Delta} \begin{bmatrix} s_0^{-1} s_1 (G - \mu_1 I) & s_0^{-1} s_1^{-1} I \\ -s_0 s_1 (-\mu_0 G - \mu_1 H + \mu_0 \mu_1 I + J) & s_0 s_1^{-1} (-H + \mu_0 I) \end{bmatrix} \tag{22}$$

where $\mu_0$, $\mu_1$ correspond to the values taken by the relative derivative: $\mu_{0,1} = s'_{0,1} / s_{0,1}$ at both sides of the layer. The terms $G, H, I, J, \Delta$ involve the values taken at the two layer edges by the independent functions $K(\xi, p)$ and $P(\xi, p)$ and by their derivatives:

$$\begin{cases} G = K_0 P'_1 - K'_1 P_0 & H = K'_0 P_1 - K_1 P'_0 \\ I = K_0 P_1 - K_1 P_0 & J = K'_0 P'_1 - K'_1 P'_0 \\ \Delta = W(K, P) = K_1 P'_1 - K'_1 P_1 = K_0 P'_0 - K'_0 P_0 \end{cases} \quad (23)$$

We now build upon the previous study and provide useful expressions for the $\text{sech}(\hat{\xi})$-profile case. When taking into account the expressions of $K(\xi, p)$ and $P(\xi, p)$ in Eq. (10)-(11) we get $\Delta = -\alpha p$ and:

$$\begin{bmatrix} G \\ H \\ I \\ J \end{bmatrix} = - \begin{bmatrix} \sigma_1(\sigma_1 - \sigma_0) + p & \sigma_0 p - \sigma_1(\alpha^2 - \sigma_0 \sigma_1) \\ \sigma_0(\sigma_1 - \sigma_0) - p & \sigma_1 p - \sigma_0(\alpha^2 - \sigma_0 \sigma_1) \\ -(\sigma_1 - \sigma_0) & \alpha^2 - \sigma_0 \sigma_1 \\ (\sigma_1 - \sigma_0)(p - \sigma_0 \sigma_1) & \alpha^2 \sigma_0 \sigma_1 - (\sigma_0^2 + p)(\sigma_1^2 + p) \end{bmatrix} \times \begin{bmatrix} \alpha \cosh(\alpha \xi_1) \\ \sinh(\alpha \xi_1) \end{bmatrix} \quad (24)$$

A careful analysis of Eq. (19) shows that the four entries of a $\mathbf{M}_{\langle \varphi \rangle}$ matrix are obtained from those of the corresponding $\mathbf{M}_{\langle T \rangle}$ matrix in Eq. (22) according to:

$$\begin{bmatrix} A_{\langle \varphi \rangle} & B_{\langle \varphi \rangle} \\ C_{\langle \varphi \rangle} & D_{\langle \varphi \rangle} \end{bmatrix} = \begin{bmatrix} D_{\langle T \rangle} & p^{-1} C_{\langle T \rangle} \\ p B_{\langle T \rangle} & A_{\langle T \rangle} \end{bmatrix} \quad (25)$$

All ingredients have now been provided for the computation of the transfer matrices related to $\text{sech}(\hat{\xi})$-type profiles. The procedure for performing the calculation is sketched thereafter:

1. specify four boundary conditions on effusivity $b(\xi)$ and its derivative at left-end and right-end of the graded layer : $b_0, b_1, b'_0, b'_1$ ;
2. choose between $\langle T \rangle$-form and $\langle \varphi \rangle$-form;
3. translate the specifications on the derivatives of effusivity into specifications on the derivatives of $s(\xi)$ with reference to Eq. (8), while considering the appropriate $+1/2$ or $-1/2$ exponent;
4. identify the four parameters $A_B$ and $A_D$, $\xi_c$ and $\tau$ such that $s(\xi)$ in Eq. (7)-(8), satisfies the four boundary conditions (specified values for $s_0, s_1, s'_0, s'_1$);
5. the whole profile $b(\xi)$ can then be drawn from Eq. (8) (by using Eq. (7)); the profile in $z$-space can be calculated from the appropriate equations in Eq. (14)-(17) depending on the form ($\langle T \rangle$ or $\langle \varphi \rangle$) and the underlying hypothesis (constant $c(z)$ or constant $\lambda(z)$);
6. for a given $p$ Laplace-Fourier parameter, compute $\alpha(p)$ and the values of $\sigma(\xi)$ at both boundaries, $\sigma_0$ $\sigma_1$, from Eq. (11). Calculate the four quadrupole entries by computing the terms in Eq. (24) and substituting them into Eq. (22) and then, if appropriate, into Eq. (25).

The quadrupole matrix relevant of an association of $\text{sech}(\hat{\xi})$-type profiles is naturally obtained by multiplying the individual matrices together.

## 5 Applications

### 5.1 Simple examples: graded coatings

Fig. 1 is an illustration of $\text{sech}(\hat{\xi})$-type profiles that adds to those already presented in [26, 28]. Those presented here are intended to represent a graded coating. Both $\langle T \rangle$-form and $\langle \varphi \rangle$-form profiles are reported (in plain, resp. dashed curves). We hypothesized a doubling in effusivity from $b_0$, at the left-end of the coating (free-surface), to $b_1$ at the right-end. We set a zero derivative at the right-end whereas the derivative was given the values 2, -1.5, or 0 (normalized derivative $\xi_1 (b')_0 / b_0$) at the left-end. Respectively, the effusivity profiles show an overshoot, an undershoot or they rise monotonically to the right-end value. In the latter case, the $\langle T \rangle$-form and $\langle \varphi \rangle$-form profiles are very similar, whereas in the former cases they are quite different, which highlights the interest in considering both $\langle T \rangle$-form and $\langle \varphi \rangle$-form options: this enriches quite simply the family of solvable profiles and gives the opportunity to choose from many more forms in the modelling process.

The representation of these $\text{sech}(\hat{\xi})$-type profiles in the (real) depth space is given in Fig. 2. Two hypotheses were considered: constant volumetric heat capacity (left plot) and constant heat conductivity (right plot). Notice that the case of constant diffusivity is of course already represented in Fig. 1 since $z$ and $\xi$ are then simply proportional. In the left plot of Fig. 2 one can observe a stretching or a compression of the initial $\xi$-profile in Fig. 1 in those places where it shows high, resp. low effusivity values. In the right plot, we observe the opposite. We then assumed that these coatings are laid over a substrate (semi-infinite layer) of effusivity $b_1$, i.e. the one at the right-edge of the coating. Thus, the effusivity is continuous up to the first derivative at the interface. The temperature response of the free surface of the coating when it is submitted to a modulated heat input of power density (amplitude) $P$ and frequency $f$ is represented in Fig. 3 (amplitude in left plot, phase in right plot). In the absence of heat losses, the front-surface temperature is given by:

$$\theta = P \frac{AZ + B}{CZ + D} \tag{26}$$

where $Z = (b_1 \sqrt{p})^{-1} = (b_1 \sqrt{i2\pi f})^{-1}$ is the thermal impedance of the homogeneous semi-infinite layer and $A, B, C, D$ are the four entries of the coating quadrupole (while taking appropriately $\langle T \rangle$-form or $\langle \varphi \rangle$-form related expressions in Eq. (22)-(25)). The normalized amplitude reported in Fig. 3-left is $|\theta b_0 / P \xi_1|$. The green curves in both plots describe the behaviour of a single homogeneous material of effusivity $b_0$ (reference case).

Amplitude and phase are plotted vs. the non-dimensional frequency $f\xi_1^2$ (values in the range of 1 indicate a frequency match with the coating diffusion time). The results in fig. 3 were validated against the classical staircase profile model. As expected, the latter results converge to the former ones when increasing the number of homogeneous sublayers (50 sublayers are necessary for getting less than 1.8% error in amplitude and 0.7° error in phase over the considered frequency range; with 100 sublayers these errors reduce to 0.4%, resp. 0.2°). In Fig. 3, as expected, with increasing frequency, the asymptotic amplitude trend and phase level are those of the reference case. Reciprocally, under vanishing frequency, they come close to those of the bulk material. In the intermediate regime, i.e. $f\xi_1^2$ between 0.01 and 100 (which is a quite wide range), notwithstanding the global $f^{-1/2}$ trend, the amplitude in Fig. 3-left follows an evolution that very roughly corresponds to the inverse of the effusivity profile. On the other hand, in Fig. 3-right, the phase evolution looks more complicated and its variations are less easy to interpret with respect to the known in-depth variations of effusivity. At first glance, one might assume that the phase contrast to the reference value -45° is loosely related with the derivative of the effusivity profile against $\xi$.

Ultimately, the three groups of profiles show extremely different thermal behaviours, suggesting that the inversion of amplitude and phase data to characterize the graded profiles is likely to succeed. However, discriminating between the presently considered $\langle T \rangle$-form and $\langle \varphi \rangle$-form profiles will be unequally easy. The highest difficulty will be with the monotonically rising profiles (blue curves) since those are anyway very close. Another way of representing the temperature amplitude $|\theta(f)|$ consists in multiplying it with $f^{1/2}$ to compensate for the general $f^{-1/2}$ trend and then taking the inverse. Indeed, computing $P/\left(|\theta(f)|\sqrt{2\pi f}\right)$ yields a frequency-dependent function that has the dimension of effusivity and whose asymptotic values are $b_0$ for $f \to \infty$ and $b(\xi \to \infty) = b_1$ for $f \to 0$. We may call this function "apparent effusivity": $b_{app}(f)$. Actually, the time-domain version of this "apparent effusivity" (i.e. as inferred from the pulse temperature response) was first presented in [31, 32] and it later found many applications in the field of photothermal measurements (see e.g. [33-35]).

In the frequency domain, a frequency scan from high to low values yields "apparent effusivity" values that are expected to provide information on the *actual effusivity* values at progressively deeper levels. That $b_{app}(f)$ dispenses a perfect image of the effusivity profile $b(\xi)$, through to a still unknown relationship between $f$ and SRDT $\xi$, is however a naive thought. Fig. 4 presents the (normalized by $b_1$) apparent effusivity curves related to all six $\text{sech}(\hat{\xi})$ profiles in Fig. 1. First, the variations of the apparent effusivity are damped with respect to those of the true profiles (non-monotonic cases). The damping is particularly important for the profiles exhibiting an overshoot (red curves). Secondly, the relation between the $\xi/\xi_1$-scale (in Fig. 1) and the frequency-scale $f\xi_1^2$ (in Fig. 4) is far from obvious; anyway it cannot be reduced to a mere $x \to 1/\sqrt{x}$ transformation. Therefore, the apparent effusivity curves should only be considered as a rough description of the actual effusivity profiles. More involved inversion procedures should be implemented to get a valuable description of these, which will be the subject of a future paper.

5.2 Synthesis of more elaborate profiles

More complex (solvable) profiles can be synthesized by joining several $\text{sech}(\hat{\xi})$-type profiles. Thanks to their flexibility, one can manage getting synthetic profiles that are continuous at the nodes up to the first or even second derivative. Figure 5 is a collection of five such synthetic profiles incorporating up to ten $\text{sech}(\hat{\xi})$-type elements (the profiles of $\sqrt{b(\xi)}/\sqrt{b_\infty}$ are represented vertically vs. $\xi/\xi_{total}$ where $\xi_{total}$ is the SRDT of all the graded part of the profile). They are backed by an homogeneous semi-infinite layer with effusivity $b_\infty$. Intricate functions of this kind could be good candidates for modelling effusivity profiles encountered, for example, in natural or artificial soils.

The associated temperature response is obtained by first multiplying together the quadrupoles related to each $\text{sech}(\hat{\xi})$-type element and then applying Eq. (26) by substituting the entries of the resulting "multilayer" quadrupole. The adiabatic temperature response (normalized amplitude and phase) at the free surface of the five synthetic profiles in Fig. 5 is reported in Fig. 6 (it was assumed that the input power is absorbed at the upper surface). The green curves describe the behaviour of a homogeneous material of effusivity $b_\infty$ (reference case) towards which all amplitude and phase curves tend under vanishing frequency. On the other side, the amplitude at asymptotically high frequency is conditioned by the effusivity value at the upper surface (inverse relationship). For that reason, in Fig. 6-left, at a normalized frequency of 100, the amplitude obtained with the effusivity profiles n°1 (cyan), n°2 (black) and n°5 (red) (the numbers refer to the synthetic profiles described in Fig. 5 from left to right) is very similar whereas it is much higher than the one obtained with the effusivity profiles n°3 (blue) and n°4 (magenta). The amplitude variations with decreasing frequency are inversely correlated with the in-depth evolution of the respective effusivity profiles. However the deep effusivity variations have a progressively lower impact on the temperature amplitude, being hidden by the shallower variations. This would considerably hinder the reconstruction process of the deep part of the profiles, which is a well-known difficulty in thermal inversion (see e.g. [9-11, 14, 16-19, 21]).

Let us underscore that the amplitude and phase curves in Fig. 6 are the *exact* thermal responses of the effusivity profiles represented in Fig. 5. These results show that there is no difficulty anymore for getting *exact temperature results* for arbitrarily complex effusivity profiles. In this perspective, the "high-level" quadrupoles related to $\text{sech}(\hat{\xi})$-type elementary profiles can be considered as sophisticated analytical Meccano pieces (in opposition to the classical straight beams). In addition to this aim, we have sought, through the choice of the geometric forms represented, to offer a mnemonic means to remember the name of the functions that are the basis of this technique: the $\text{sech}(\hat{\xi})$ profiles ([sek ksi hæt] = [seksi hæt]). For this purpose, the synthetic profiles in Fig. 5 have been superimposed in Fig. 7.

## 6 Discussion and extension to other transfer equations (Maxwell's equations)

The previous results highlighted the advantages of exploiting the $\text{sech}(\hat{\xi})$-type profile concept for solving the *direct* thermal problem in graded media. The dynamical temperature data referred to the frequency domain. For getting, instead, the transient data corresponding to a pulsed excitation, one should still perform an inverse Laplace transformation on the temperature expression in Eq.(26) (whilst involving the Laplace variable $p$). This is easily and efficiently realized numerically by the De Hoog method [30, 36]. In the next future, the *inverse* problem will be tackled, namely the identification of an effusivity profile (through a $\text{sech}(\hat{\xi})$-type profile description) from the pulsed or modulated photothermal data measured on the outer surface of the heterogeneous material.

The PROFIDT method and the related products, among them the powerful $\text{sech}(\hat{\xi})$-type profiles, are not restricted to the heat equation. They can also be applied to wave equations used to model plane wave propagation in 1D graded media like in optics (Maxwell's equations), in acoustics (acoustic waves and shear waves) and in heterogeneous transmission lines (telegraphist equations). The adaptation of the PROFIDT method to the Maxwell's equations in lossless materials with graded permittivity and permeability was performed in [28]. Optical materials with a graded refractive-index present a subcase.

There are actually interesting equivalences with the heat transfer problem. Roughly speaking there is a close connection between on one side, the effusivity, the SRDT, $\langle T \rangle$-form and $\langle \varphi \rangle$-form profiles and on the other side the (effective) refractive index, the optical thickness, $\langle E \rangle$-form and $\langle H \rangle$-form profiles where $E$ and $H$ stand for the electric and magnetic fields. In essence, the $\langle T \rangle$-form quadrupole in Eq. (22), (24) is the same as the $\langle E \rangle$-form quadrupole; one simply needs to perform the following change $p \leftrightarrow -k_0^2$ where $k_0$ is the wavenumber in vacuum of the light wave (the quadrupole transformation from $\langle E \rangle$-form to $\langle H \rangle$-form is however not the same as between $\langle T \rangle$-form and $\langle \varphi \rangle$-form in Eq. (25); it actually reduces to a two-step circular permutation, see [28]).

The $s(\xi)$ profiles are strictly the same for both problems, which makes that the $\text{sech}(\hat{\xi})$-type profiles present the same (great) interest for both. As an example, a multiplicity of $\text{sech}(\hat{\xi})$-type profiles with zero end-slopes where joined for calculating the exact reflectance/transmittance spectra of locally-periodic refractive-index profiles encountered in rugate filters, Bragg gratings or in chirped mirrors [28].

More illustrations of the power of the $\text{sech}(\hat{\xi})$-type profiles for diffusion and wave modelling in continuously heterogeneous media, both for direct and inverse problems, are expected in the future.

**Figures**

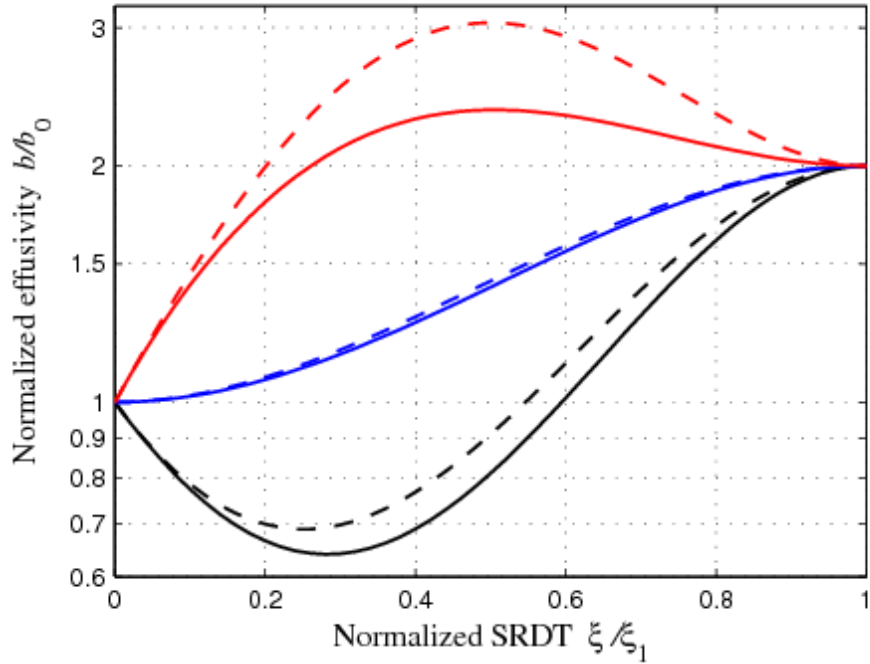

**Fig. 1** Effusivity profiles of $\text{sech}(\hat{\xi})$–type with various boundary specifications. The effusivity $b(\xi)$ is normalized by the left-edge value $b_0$. The profiles are plotted against the normalized SRDT (square root of diffusion time) $\xi/\xi_1$ where $\xi_1$ is the SRDT of the graded layer. The profiles of $\langle T \rangle$-form are in continuous lines, those of $\langle \varphi \rangle$-form are in dashed lines. The right-to-left effusivity ratio $b_1/b_0$ is 2. Three values are considered for the left-end (dimensionless) derivative $\xi_1(b')_0/b_0$ : -1.5 (black), 0 (blue) and +2 (red). The right-end derivative is 0 for all six cases.

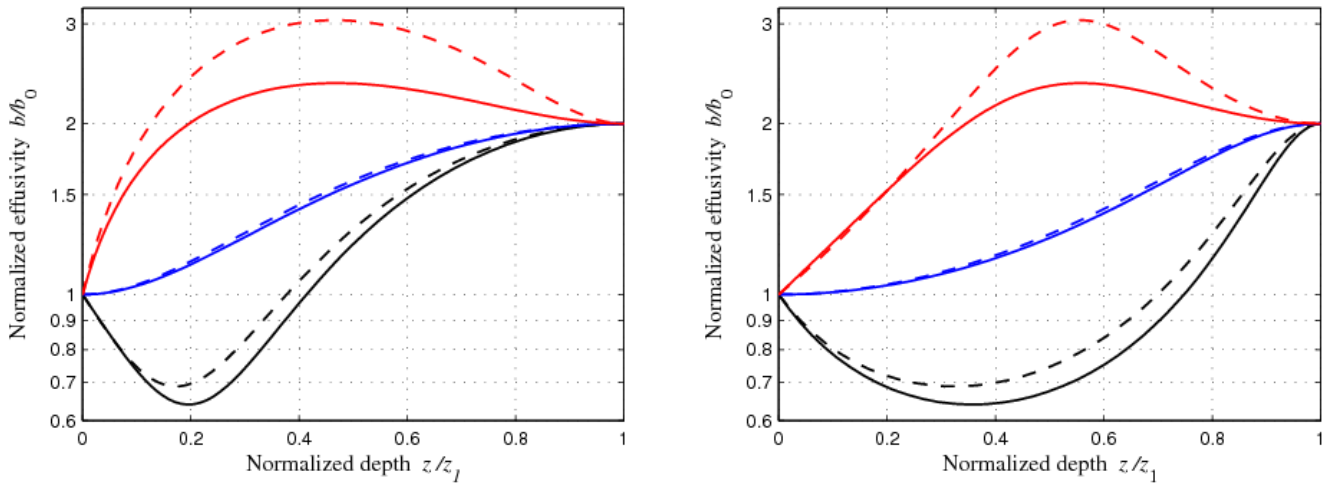

**Fig. 2** Effusivity profiles of Fig. 1 represented in the physical-depth space, i.e. vs $z/z_1$ where $z$ is distance from the left boundary and $z_1$ is the layer thickness. Two cases were considered for the $\xi \to z$ (inverse-Liouville) transformation. In the left plot, the underlying assumption is that volumetric heat capacity is constant in the graded layer. In the right plot, heat conductivity was assumed constant.

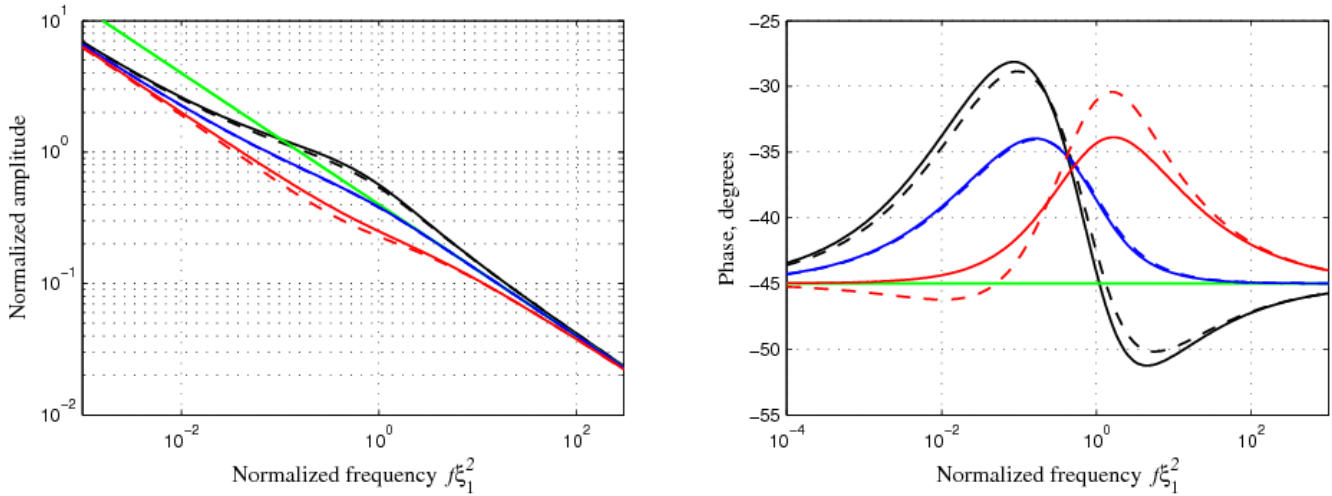

**Fig. 3** Temperature response (left is the normalized amplitude –see text– and right is the phase) of the profiles drawn in Fig. 1 and Fig. 2 when laid over a semi-infinite layer of effusivity $b_1$ and submitted at the left boundary $\xi = 0$ to a modulated heat input of frequency $f$. The amplitude and phase are plotted vs the normalized frequency $f\xi_1^2$. Same colours as in Fig. 1 and Fig. 2. The additional green line corresponds to the response of a homogeneous semi-infinite material whose effusivity is equal to the surface effusivity of the graded profiles, $b_0$.

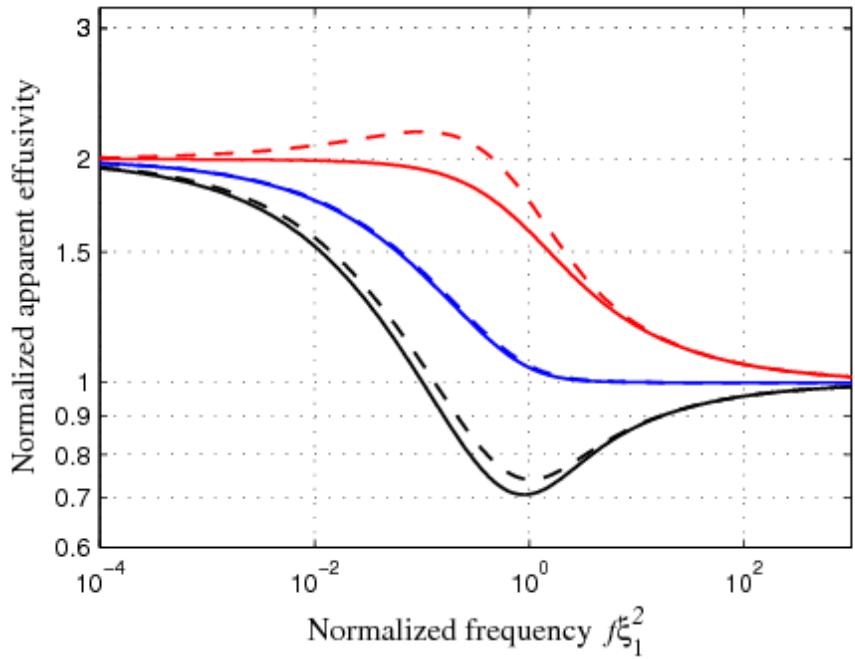

**Fig. 4** Same as in fig. 3 for the (normalized) apparent effusivity which corresponds to the ratio between the amplitude of the homogeneous (reference) material and the amplitude of the graded material. This apparent effusivity is a frequency function; it is here plotted vs. the normalized frequency $f\xi_1^2$.

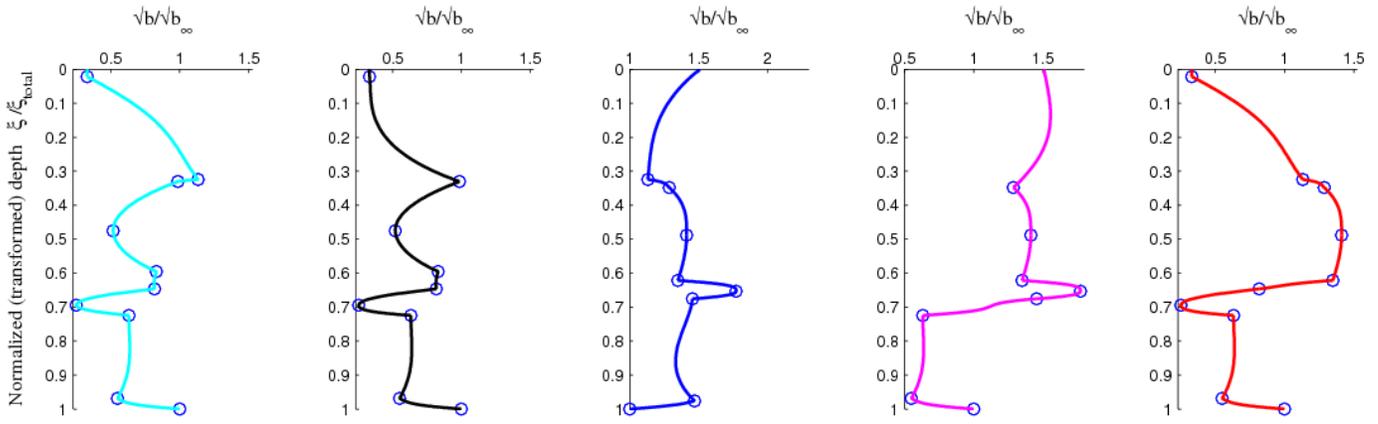

**Fig. 5** A series of five composite effusivity profiles as obtained by joining up to ten $\text{sech}(\hat{\xi})$–type profiles. The vertical profiles of the square root of effusivity (normalized by the bulk value) is plotted vs. the normalized SRDT (square root of diffusion time) $\xi/\xi_{total}$ where $\xi_{total}$ is the SRDT of the whole graded profile. Blue circles indicate the connection nodes.

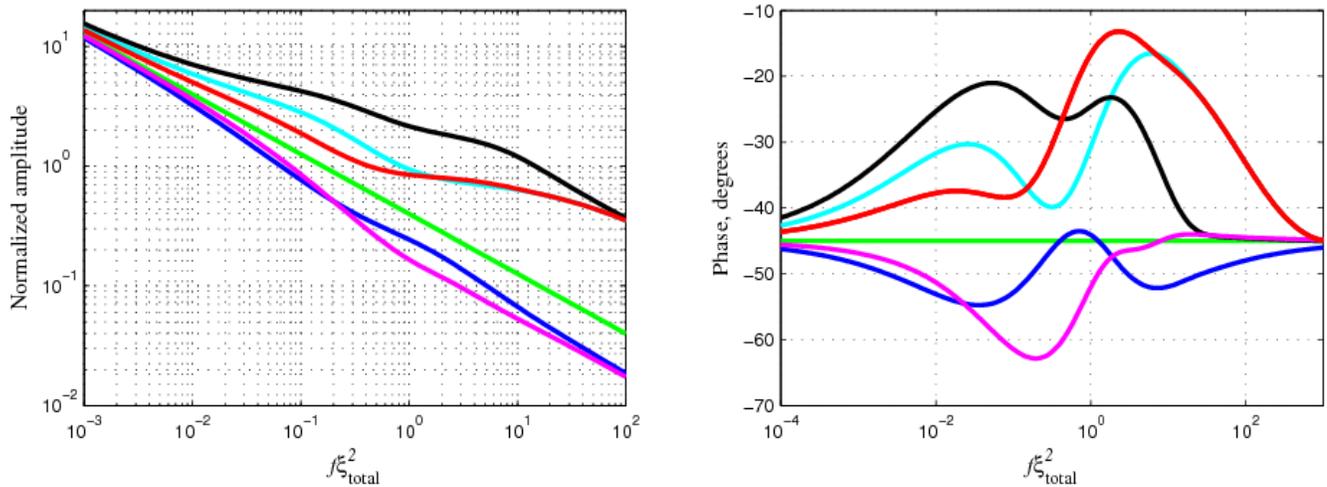

**Fig. 6** Temperature response (left : normalized amplitude, right : phase) of the profiles drawn in Fig. 5 while laid over a semi-infinite layer of effusivity $b_\infty$ and submitted at the top boundary $\xi = 0$ to a modulated heat input of frequency $f$. The amplitude and phase are plotted vs the normalized frequency $f\xi_{tot}^2$ (for the colours, refer to Fig. 5). The additional green lines correspond to the response of a homogeneous semi-infinite material whose effusivity is equal to the bulk effusivity of the graded profiles, i.e. $b_\infty$.

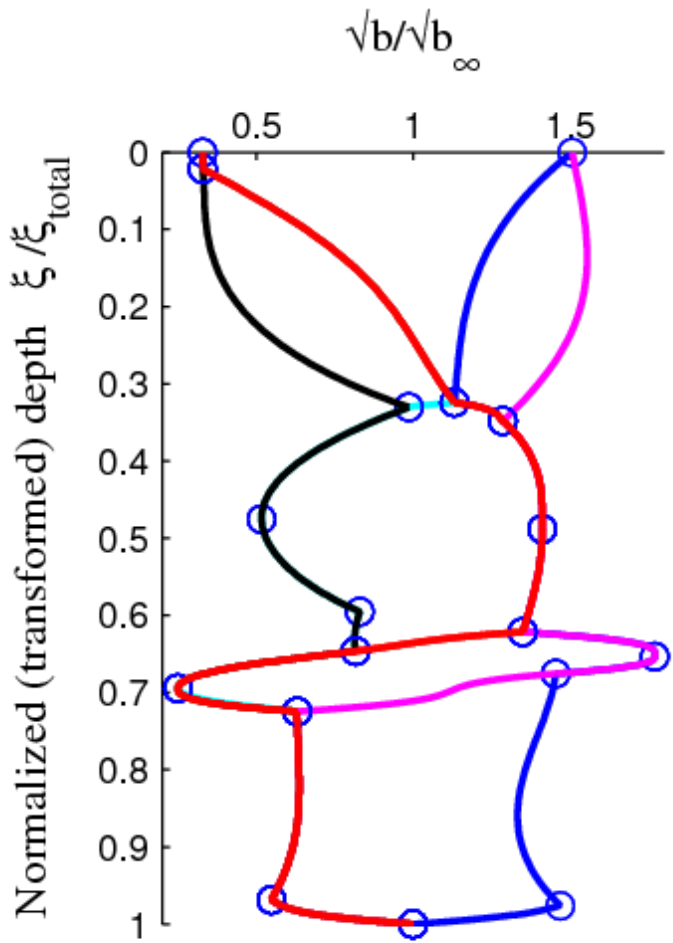

**Fig. 7** Superposition of the five synthetic profiles from Fig. 5. In addition to its mnemonic power, this figure intends to show the large variety of shapes one can build from the basis of $\text{sech}(\hat{\xi})$ ([seksi hæt]) profiles.